\documentclass[%
 preprint, 
 amsmath,amssymb,
 aps, physrev,
]{revtex4-2}

\usepackage{graphicx}
\usepackage{dcolumn}
\usepackage{bm}

\begin{document}


\title{\textbf{Direct observation of time-dependent coherent chiral tunneling dynamics.} 
}%

\author{Wenhao Sun}
\author{Denis S. Tikhonov}%
\altaffiliation{Current address: Center for Free-Electron Laser Science CFEL, Deutsches Elektronen-Synchrotron DESY, Notkestr. 85, 22607 Hamburg, Germany.}

\author{Melanie Schnell}
\email{melanie.schnell@desy.de}
\altaffiliation{Institute of Physical Chemistry, Christian-Albrechts-Universit\"{a}t zu Kiel, Max-Eyth-Str. 1, 24118 Kiel, Germany}

\affiliation{
Deutsches Elektronen-Synchrotron DESY, Notkestr. 85, 22607 Hamburg, Germany
}%

\date{\today}

\begin{abstract}

Superpositions of handed molecular states give rise to achiral eigenstates, delocalized across a double-well potential via tunneling. A coherent superposition of these energy eigenstates could dynamically relocalize the molecules into chiral states, which has only been addressed theoretically. Here, we present a microwave pump-probe study to create and probe coherent chiral tunneling dynamics in a given rotational state. Through a time-resolved scheme, we uncover the periodic time evolution of the induced chiral wavepacket under field-free conditions. Moreover, we demonstrate precise phase control of this coherence via phase modulation during pump excitation.  

\end{abstract}

\maketitle

\begin{figure}[!t]
\centering
\includegraphics{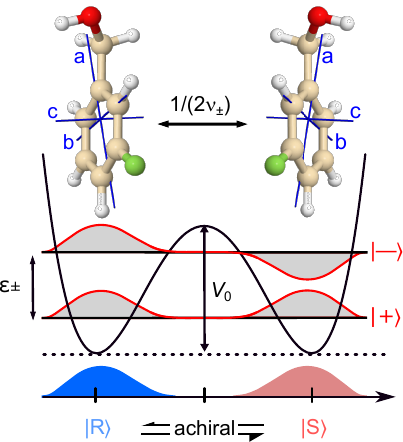}
\caption{Double-well potential of 3-fluorobenzyl alcohol. Molecular geometries of \textit{R}- and \textit{S}-3-fluorobenzyl alcohol are shown in their inertial principal axis systems (PAS). The $|R\rangle$ and $|S\rangle$ states represent the localized right- and left-handed chiral states, whereas the $|\pm\rangle$ states denote achiral energy eigenstates with opposite parity in the vibronic ground state. The energy difference $\varepsilon_\pm=h\nu_{\pm}$ between the $|\pm\rangle$ states in the $J_{K_aK_c} = 0_{00}$ state of 3-fluorobenzyl alcohol, expressed through the tunneling frequency $\nu_{\pm}$, has been experimentally determined to be $\nu_{\pm} = 0.818(12)$ MHz \cite{3fba_cpl2010}. Accordingly, the transfer time from one chiral form to the other form is $1/(2\nu_\pm)$, that is 0.611 $\mu$s.}
\label{fig:potential}
\end{figure}

\textit{Introduction.}\textemdash Molecular chirality is a fascinating phenomenon where a molecule can exist in left- and right-handed forms that are mirror images of each other, also known as enantiomers. Although enantiomers share nearly identical physical properties, their non-superimposable structures often lead to distinct chemical and biological behaviors. Despite its widespread presence in our everyday life, molecular chirality remains a field rich in unresolved mysteries, such as its origin in physics and biological homochirality \cite{quack1989,homochirality_CR2004}. In quantum theory, chiral molecules are described as two localized chiral states in a double-well system \cite{hund1927-1,hund1927-2}, as depicted in FIG. \ref{fig:potential}. The two potential wells are identical, up to a possible very small energy difference ($\Delta E_\mathrm{PV}$) due
to the parity-violation effect \cite{pv_angew2002}. Governed by the double-well potential, molecules may exist in non-stationary chiral configurations localized in one of the wells or in stationary achiral states delocalized over both wells in equal amounts \cite{homochirality_csr1988,h2o2_cp2007}. The latter corresponds to a pair of energy eigenstates with symmetric and antisymmetric parity ($|\pm\rangle$) under spatial inversion, resulting from the tunneling of chiral states through the potential barrier \cite{hund1927-1,hund1927-2,quack1989}. Nevertheless, the boundary between the classical-like chiral states and quantum-mechanical superposition states are still unclear, which is commonly referred to as Hund's paradox \cite{hund1927-2,primas2013}.

Addressing such problems requires high-resolution spectroscopy and precise quantum control of molecular systems using external fields. Based on the quantum superposition principle, two stable enantiomers (\textit{R} and \textit{S}) can be prepared as a quantum superposition with well-defined parity ($|\pm\rangle$) through subtle matter-wave interactions, generating an observable tunneling dynamics \cite{superposition_jcp1994,superposition_science1995,daqing2021}. In turn, coherent superposition of the vibrational-tunneling $|\pm\rangle$ eigenstates will lead to coherent chiral tunneling within the double-well potential \cite{ammonia_jcp2019,ct_jcp2020,cdt_prl1991}. This gives rise to a non-stationary chiral wavepacket, periodically oscillating at the tunneling frequency under field-free conditions. However, it still remains a challenging task to experimentally establish the quantum conversion between chiral states and delocalized superposition eigenstates \cite{ct_nature2010}.

Over the past decade, the microwave three-wave mixing (M3WM) technique, a resonant, coherent, and non-linear approach, has demonstrated its robust capability in preparing coherent states for stable chiral molecules with enantiomeric sensitivity \cite{m3wm_PJASB2012,m3wm_Nature2013,esst_prl2017,chm_jpcl2018}. This technique was later adapted to create a coherent superposition of energy eigenstates within the rotational states of transiently chiral molecules \cite{m6wm_nc2023}. In an earlier experiment using benzyl alcohol as a model, an enantiomeric excess was induced from the quantum racemic mixture, suggesting the successful preparation of the coherent tunneling state \cite{m6wm_nc2023}. However, a direct observation of its time-dependent dynamics was not achieved due to its rate being faster than the capabilities of the experimental setup.

Herein, we demonstrated a pump-probe experiment using a quantum racemic mixture of 3-fluorobenzyl alcohol, revealing the field-free time evolution of the non-stationary chiral wavepacket arising from coherent tunneling. The experiment begins with a pump cycle that prepares a coherent superposition of the $|\pm\rangle$ eigenstates in a chosen rotational level, driven by a combination of two-photon and one-photon excitations. Although the resulting dynamically localized chiral wavepacket is non-observable due to the forbidden nature of the $|+\rangle \leftrightarrow |-\rangle$ transition, it produces a macroscopic enantiomeric excess, which can be detected by a subsequent probe cycle. Two probing sub-cycles are simultaneously driven to induce polarizations at two listen transitions. The observed time-domain listen signal exhibits interference beats in the form of a free induction decay (FID). By varying the time delay between the pump and probe cycles, we capture the time evolution of these beats under field-free conditions, offering direct observation of the time-dependent coherent chiral tunneling dynamics in the target state. Additionally, we demonstrate that phase modulation during the pump excitation allows precise control over the phase of the induced coherent tunneling.

\begin{figure}[!t]
\centering
\includegraphics[width=\linewidth]{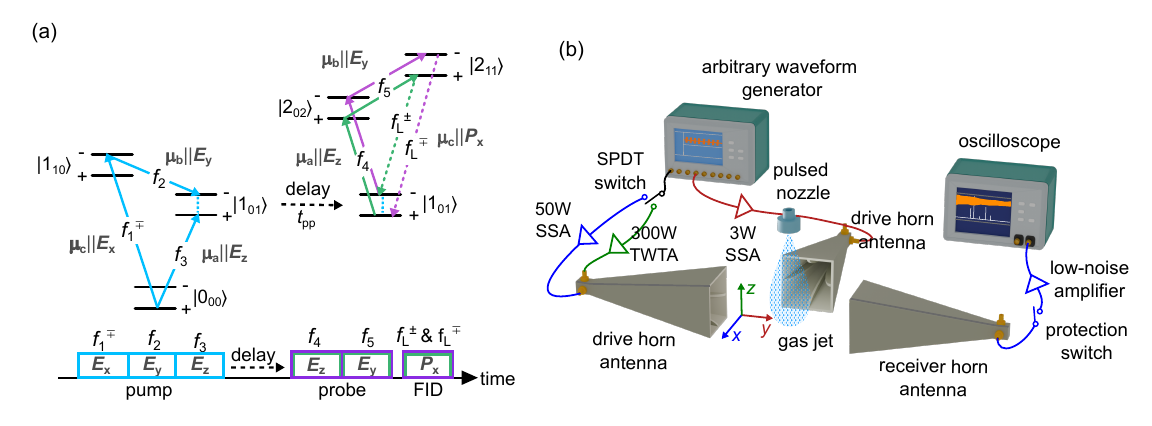}
\caption{Level scheme and experimental setup for the pump-probe measurement. (a). Pump-probe level diagram starting from the $|{0_{00}}^{+}\rangle$ rotational state, with the pulse sequence shown below. The energy levels are marked with the rotational state $|J_{K_aK_c}\rangle$ and the superposition parity $\pm$. The \textit{M}-degeneracy of rotational levels is omitted for clarity. Intrastate transitions are denoted using $f_\mathrm{i}$, where tunneling doublets can be excited using the same frequency, while the interstate transition components are labeled explicitly with ${f_\mathrm{i}}^{\mp}$/${f_\mathrm{i}}^{\pm}$. For each transition, the corresponding dipole-moment component in the molecular frame and the polarization direction of the microwave field in the laboratory frame are indicated using $\mathbf{\boldsymbol{\mu}_{\alpha}} || \mathbf{E}_{\beta}$ ($\alpha=a,b,c$; $\beta=x,y,z$). In both pump and probe cycles, excitations of the three types of transitions take place in mutually orthogonal polarization directions. (b). Schematic of the Fourier transform microwave spectrometer COMPACT used in this work. See Supplementary Section II B for detailed demonstration and abbreviations \cite{SM}. }
\label{fig:setup}
\end{figure}

\textit{Theory.}\textemdash As can be seen in FIG. \ref{fig:potential}, the tunneling motion in 3-fluorobenzyl alcohol arises from a concerted rotation of the CH$_2$OH and OH groups, transferring one enantiomer into the other enantiomeric form. This large-amplitude motion (LAM) is hindered by a barrier of 1.9 kJ/mol, resulting in a tunneling frequency of $\nu_\pm =$ 0.818(12) MHz \cite{3fba_cpl2010}. Thus, 3-fluorobenzyl alcohol exists in the achiral superposition eigenstates, whereas the right- and left-handed chiral states are non-stationary. Due to the selection rules, direct excitation between the eigenstates of opposite parity ($|+\rangle \leftrightarrow |-\rangle$) is dipole-forbidden within the same rotational level. But even if possible, a one-photon excitation cannot create a chiral ensemble, but only a racemic mixture \cite{TikhonovSA2022}. To create a macroscopic ensemble of coherently superposed $|\pm\rangle$ states, a microwave excitation scheme, adapted from the M3WM approach, is proposed, as presented in FIG. \ref{fig:setup}a.

In the level diagram, each state is labeled with the rotational level $|J_{K_aK_c}\rangle$ and the associated parity $\pm$, with $J$ being the total angular momentum quantum number, and $K_a$ and $K_c$ being projections of this momentum onto the $a$- and $c$-principal axes in the molecular frame. As the tunneling motion inverts the electric dipole-moment component along the $c$-principal axis, $\mu_c$-type rotational transitions ($f_1$ and $f_\mathrm{L}$) are allowed between tunneling substates of opposite parity ($f^{\mp}$: $|+\rangle \rightarrow |-\rangle$ and $f^{\pm}$: $|-\rangle \rightarrow |+\rangle$), which are called interstate transitions. The frequency difference of the two interstate transition components is approximately $\Delta \nu = 2\varepsilon_\pm/h$, which is 1.64 MHz in this system. On the other hand, $\mu_a$- and $\mu_b$-type transitions ($f_2-f_5$) are only allowed for intrastate transitions ($|+\rangle \rightarrow |+\rangle$ and $|-\rangle \rightarrow |-\rangle$). The splittings of intrastate transitions are small for 3-fluorobenzyl alcohol, so each tunneling doublet can be simultaneously excited using the same frequency.

Starting from a $\mu_c$-type transition ${f_\mathrm{1}}^{\mp}$, molecules initially populating the $|{0_{00}}^+\rangle$ state are excited to $|{1_{01}}^-\rangle$ and $|{1_{01}}^+\rangle$ states, following the two-photon ($|{0_{00}}^{+}\rangle \xrightarrow{{f_1}^{\mp}} |{1_{10}}^{-}\rangle \xrightarrow{f_2}  |{1_{01}}^{-}\rangle$) and one-photon ($|{0_{00}}^{+}\rangle \xrightarrow{f_3}  |{1_{01}}^{+}\rangle$) pathways (FIG. \ref{fig:setup}a), respectively. The two-path excitation creates a coherent superposition of the $|\pm\rangle$ eigenstates in the $|{1_{01}}\rangle$ rotational level, even though one-photon excitation is strictly forbidden. Disregarding the absolute phase of this superposition, the initial state of the induced coherence in $|{1_{01}}\rangle$ can be described as
\begin{equation}
    \Psi(t=0) = \frac{1}{\sqrt{2}}\left(|{1_{01}}^+\rangle + \exp(i\phi) |{1_{01}}^-\rangle\right),
\end{equation}
where $|{1_{01}}^\pm\rangle = \frac{1}{\sqrt{2}}(|R_{1_{01}}\rangle \pm |S_{1_{01}}\rangle$) with $|R_{1_{01}}\rangle$ and $|S_{1_{01}}\rangle$ being the non-stationary wavefunctions of $R$ and $S$ enantiomer in the $|{1_{01}}\rangle$ state and $\phi$ is the pulse-controlled initial phase of the wavepacket. This initial state corresponds to a chiral wavepacket localized in one of the potential wells, which subsequently undergoes field-free time evolution governed by the LAM tunneling motion. Again, as the $|{1_{01}}^-\rangle \leftrightarrow |{1_{01}}^+\rangle$ transition is dipole-forbidden, this tunneling coherence cannot be spectroscopically probed.

Instead, the chiral wavepacket produces a macroscopic enantiomeric excess (\textit{ee}) in the $|1_{01}\rangle$ state \cite{m6wm_nc2023}. The resulting \textit{ee} follows the same time-dependent dynamics as the non-stationary chiral wavepacket,
\begin{equation}
 \label{eq:eeplus}
\langle \psi(t)|  \hat{ee}_{1_{01}} |\psi(t) \rangle \propto
 \overbrace{- \sin(\Omega_1 \tau_1) \sin\left( \frac{\Omega_2 \tau_2}{2} \right)
 \sin\left( \frac{\Omega_3 \tau_3}{2} \right)}^{\propto |ee|} \cdot
 \sin\left(2\pi\nu_{\pm}t +\varphi_1 - \varphi_2 - \varphi_3 + \Phi \right) \ ,
\end{equation}
where the $\hat{ee}$ operator for the $|1_{01}\rangle$ state can be expressed as \cite{m6wm_nc2023} 
\begin{equation}
 \label{eq:eedef}    
\hat{ee}_{1_{01}} = |R_{1_{01}}\rangle \langle R_{1_{01}} | - |S_{1_{01}}\rangle \langle S_{1_{01}}| = |{1_{01}}^+\rangle \langle {1_{01}}^-| + |{1_{01}}^-\rangle \langle {1_{01}}^+| \ , 
\end{equation}
$\Omega_k$, $\tau_k$ and $\varphi_k$ denote the Rabi frequency, duration, and starting phase of the \textit{k}-th microwave pulse in the scheme, $\Phi$ represents the accumulated phase throughout the excitation process, $\nu_\pm$ is the tunneling frequency in the $|1_{01}\rangle$ state, and $t$ is the time.

As described in Equation \ref{eq:eeplus}, this \textit{ee} observable oscillates at the tunneling frequency ($\nu_\pm$) under field-free conditions, with a constant amplitude proportional to $|ee|$. The time evolution of \textit{ee} can thus be monitored through a chiral-sensitive probe cycle, in which two microwave excitations ($f_4$ and $f_5$) and an indirectly induced listen transition (${f_\mathrm{L}}^{\pm}$ or ${f_\mathrm{L}}^{\mp}$), combined with the tunneling coherence create a closed loop. In particular, as $f_4$ and $f_5$ are associated with intrastate transitions, two probing sub-cycles can be simultaneously driven, sensing the same tunneling coherence from opposite directions. At the two $\mu_c$-type interstate listen transitions, the signal phases should be linearly dependent on the tunneling coherence oscillation ($2\pi\nu_\pm t$), while the amplitudes are proportional to $|ee|$. The detailed theoretical derivations are provided in Supplementary Section I B \cite{SM}. In addition, following the pump-probe excitation, the entire process exhibits exponential decay, mainly driven by collisional relaxation and the Doppler effect, though not explicitly demonstrated for clarity.

\textit{Experiment}\textemdash The experiments are performed with an adapted broadband chirped-pulse Fourier transform microwave (FTMW) spectrometer, as shown in FIG. \ref{fig:setup}b. The principles and the modifications of the spectrometer have been described in detail elsewhere \cite{compact,m6wm_nc2023}. A brief description is provided here. The spectrometer is equipped with a two-channel arbitrary waveform generator (AWG) with a maximum sampling rate of 24 GSa/s, which is capable of generating microwave pulses below 6 GHz in a two-channel mode. Microwave pulses produced from the same AWG channel can be further divided by a single-pole double-throw (SPDT) PIN diode switch with nanosecond switching capability. In this way, microwave pulses can be separately fed into dedicated amplifiers for power amplification and broadcast into the spectrometer chamber in designated polarization directions ($\mathbf{E_x, E_y, E_z}$). The latter is achieved through two dual-polarization horn antennae oriented perpendicularly to each other.

In this study, 3-fluorobenzyl alcohol is employed as the molecular model. The sample is commercially available from Fisher Scientific with a chemical purity of 98 \% and used without further purification. The liquid sample is held in a home-built heatable reservoir, which is integrated into the solenoid nozzle valve (General Valve Series 9) and maintained at 100 \textdegree C. The diameter of the orifice is 1 mm. The sample vapor is diluted into neon carrier gas at a backing pressure of 3~bar and supersonically expanded into the vacuum chamber perpendicular to the propagation directions of the microwave radiation. The background pressure in the vacuum chamber is approximately 10$^{-6}$ mbar, and the solenoid valve is operated at 8 Hz. Molecules in the jet expansion are cooled down to a rotational temperature ($T_\mathrm{rot}$) of about 1 K. In the interaction region, the transmitted microwave pulse sequence consecutively polarizes these molecules eight times. Following each interaction sequence, the decay of the macroscopic polarization of the molecular ensemble is collected by a dual-polarization receiver horn facing one of the drive horns. The free induction decay (FID) signals are recorded with a duration of 40 $\mu$s and averaged on a fast digital oscilloscope, which is used for further data processing. This provides sufficient time resolution to unveil the tunneling dynamics of 3-fluorobenzyl alcohol.    

\begin{figure}[!b]
\centering
\includegraphics[width=0.9\linewidth]{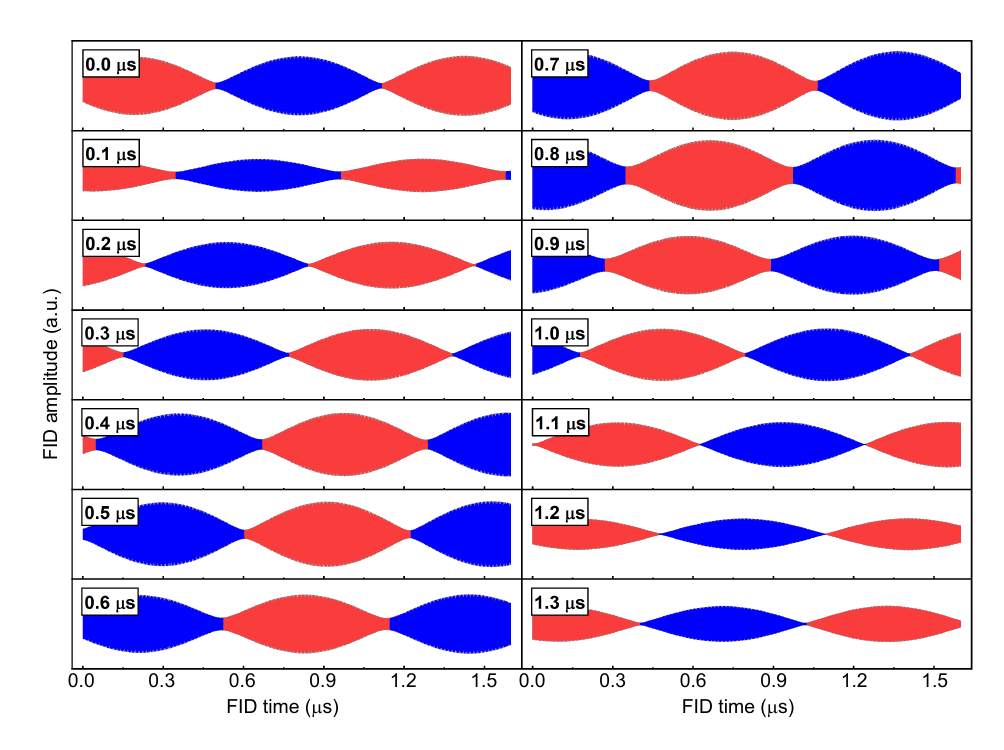}
\caption{Time evolution of the beating listen signals. The first 1.6 $\mu$s of the filtered experimental FID signals are shown at different pump-probe delays, with the delay time ($t_\mathrm{pp}$) indicated in the upper left corner. The y axis ranges from -4 to 4 nV. The original data are filtered at the listen frequencies (${f_\mathrm{L}}^{\pm}$: 6385.55 MHz and ${f_\mathrm{L}}^{\mp}$: 6387.18 MHz) using a bandpass Butterworth filter (6th-order) with a bandwidth of 60 kHz. The neighboring beats are highlighted in red and blue to indicate the sign change of the cosine-shaped beat envelopes. Note that the carrier frequency of the beats is 6385.365 MHz, which is well captured on the fast oscilloscope at a sampling rate of 25 GSa/s but too fast to be visible here due to the plotted time scale.}
\label{fig:fids}
\end{figure}

\textit{Results.}\textemdash Prior to the pump-probe experiments, the excitation microwave pulses are individually optimized by measuring Rabi cycles as a function of pulse duration. More details can be found in Supplementary Section II C \cite{SM}. In general, a Rabi $\pi/2$ pulse can evenly distribute the population in a two-level system, achieving maximum coherence between the states. In contrast, a $\pi$ pulse inverts the population between the two levels, which can be applied to transfer prepared coherence in a three-level system \cite{m3wm_jpcl2015}. Ideally, ${f_1}^{\mp}$ and $f_4$ should be $\pi$/2 pulses, whereas $f_2$, $f_3$, and $f_5$ should be $\pi$ pulses. However, in this scheme, ${f_1}^{\mp}$ and $f_3$ are optimized to near $\pi$/2 conditions, to minimize the off-resonance excitation of the ${f_1}^{\pm}$ transition and the population inversion between the $|{1_{01}}^-\rangle$ and $|{0_{00}}^-\rangle$ states, respectively. The applied pulse conditions are summarized in Supplementary Table S2 \cite{SM}.

Accordingly, the microwave pulse sequence is constructed for the pump-probe experiments, as shown in FIG. \ref{fig:setup}a. As the tunneling period of 3-fluorobenzyl alcohol in state $|{1_{01}}\rangle$ is approximately 1.2 $\mu$s, the entire probe cycle is shifted in our experiments by 0--1.3 $\mu$s in 0.1 $\mu$s steps relative to the pump cycle, while the durations and phases of all microwave pulses remain fixed. FIG. \ref{fig:fids} displays the first 1.6 $\mu$s of the time-domain FID signals recorded at each delay time ($t_\mathrm{pp}$). Owing to the similar frequencies of ${f_\mathrm{L}}^{\pm}$ and ${f_\mathrm{L}}^{\mp}$, the FID traces exhibit clear interference beats, with a carrier-wave frequency of (${f_\mathrm{L}}^{\pm} + {f_\mathrm{L}}^{\mp}$)/2 and a beat frequency of (${f_\mathrm{L}}^{\pm} - {f_\mathrm{L}}^{\mp}$)/2. A complete cycle of the interference consists of two beats, each arising from an enantiomeric excess of opposite handedness, respectively. For clarity, these beats are highlighted in different colors. As the delay time varies, the beating signals exhibit clear shifts, highlighting the non-stationary nature of the localized wavepacket in the $|{1_{01}}\rangle$ rotational level. This periodic time evolution shows good consistency with the tunneling period of 3-fluorobenzyl alcohol.      

\begin{figure}[!b]
\centering
\includegraphics[width=\linewidth]{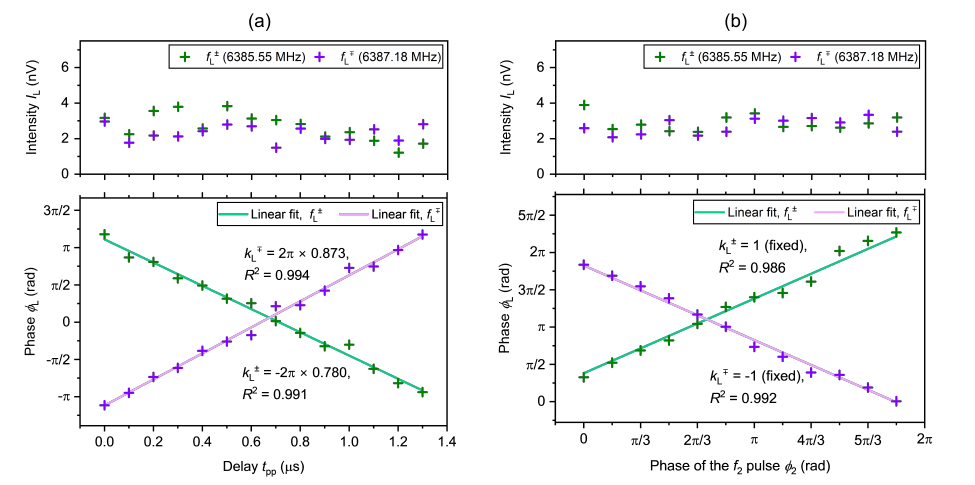}
\caption{Results of the microwave pump-probe experiments. Panels A and B show the intensities ($I_\mathrm{L}$) and phases ($\phi_\mathrm{L}$) at the listen transitions, ${f_\mathrm{L}}^{\pm}$ (6385.55 MHz) and ${f_\mathrm{L}}^{\mp}$ (6387.18 MHz), as a function of time delay ($t_{pp}$, in $\mu$s) and phase of the $f_2$ pulse ($\phi_2$, in radians), respectively. The scatter points display the experimental results. Each point is averaged from 5$\times 10^{5}$ FID acquisitions, which takes about 2.2 h. The colored lines represent the linear fits on the phase results of ${f_\mathrm{L}}^{\pm}$ and ${f_\mathrm{L}}^{\mp}$, with the slopes ($k_\mathrm{L}$) and R-squared values ($R^{2}$) provided.}
\label{fig:fft_result}
\end{figure}

After applying a fast Fourier transformation, the obtained intensities and phases at the listen transitions are presented in FIG. \ref{fig:fft_result}a. Considering the signal fluctuation over the long measurement time and the interference from a counterpart pump cycle starting from $|{0_{00}}^-\rangle$, the observed intensities indicate that the amplitudes of the listen transitions are relatively stable. The corresponding phases exhibit linear correlations with the delay time --- negative for ${f_\mathrm{L}}^{\pm}$ and positive for ${f_\mathrm{L}}^{\mp}$, as the two sub-cycles probe the tunneling coherence from opposite directions. Both the intensity and phase behaviors are consistent with the theoretical model.

Linear least-squares fits are performed on the phase results of ${f_\mathrm{L}}^{\pm}$ and ${f_\mathrm{L}}^{\mp}$, respectively (FIG. \ref{fig:fft_result}a). For both fits, the R-squared values ($R^{2}$) are higher than 0.99, suggesting nearly perfect linear correlations between the phase evolution and the time delay. The absolute values of the slopes, $|{k_\mathrm{L}}^{\pm}|$ and $|{k_\mathrm{L}}^{\mp}|$, represent the tunneling frequency in radians (2$\pi\nu_\pm$) detected by each probing sub-cycle. The two fitted tunneling frequencies deviate almost symmetrically from the spectroscopically determined value ($2\pi \times 0.82$ MHz) by approximately 5\%. This deviation can be attributed to the perturbation of the other cycle starting from the $|{0_{00}}^-\rangle$ state, similar to what was observed in Ref.~\citenum{esst_angew2023}. A detailed demonstration is provided in Supplementary Section II D \cite{SM}. By averaging these two slopes, the tunneling frequency from this time-resolved pump-probe experiment is estimated to be 0.827(15) MHz, precisely matching the spectroscopically determined values of 0.82 MHz, which falls in the frequency accuracy (10 kHz) of our spectrometer.

Furthermore, the phase of the coherent superposition of the $|\pm\rangle$ eigenstates can be finely modulated using the phases of the three pump pulses ($\varphi_1$, $\varphi_2$ and $\varphi_3$). This allows for precise control over the positioning of the induced chiral wavepacket ($\Psi(t)$) within the double-well potential. To examine this, a phase-controlled pump-probe experiment is conducted, where the phase of the $f_2$ pulse ($\varphi_2$) is scanned from 0 to 2$\pi$ in steps of $\pi$/6 radians, without applying a delay time between the pump and probe cycles. According to Equation \ref{eq:eeplus}, constant amplitudes and linear $\varphi_2$-phase dependence are expected for the listen transitions (${f_\mathrm{L}}^{\pm}$ and ${f_\mathrm{L}}^{\mp}$), similar to the time-resolved pump-probe experiment (FIG. \ref{fig:fft_result}a). The experimental results are provided in FIG. \ref{fig:fft_result}b, in convincing agreement with the expectations. The phases are linearly fitted with the slope constrained to $\pm$1. The $R^2$ values, close to 0.99, confirm the linear phase-phase correlation. As the $\varphi_2$ and $2\pi\nu_\pm t$ terms in Equation \ref{eq:eeplus} have opposite signs, the observed phase-phase and phase-delay correlations exhibit opposite trends for the same listen transition. By utilizing phase control, the chiral wavepacket can be readily prepared at various coherence phases for further advanced experiments, such as enantiomer-selective excitation, without the need to adjust experimental timing.


\textit{Discussion.}\textemdash Time-dependent coherent tunneling dynamics had, until now, solely been addressed theoretically. Our proof-of-concept experiment offers direct evidence to the theory model, exemplifying wave-particle duality and the superposition principle at the molecular level. Compared to the previously studied benzyl alcohol, which has a tunneling frequency of about 493 MHz \cite{m6wm_nc2023}, the relatively slow tunneling motion in 3-fluorobenzyl alcohol with a frequency of 0.8~MHz offers two key advantages. First, it enables the simultaneous excitation of two probing sub-cycles, producing a listen signal in the form of frequency beats. Second, the sub-MHz tunneling motion is well-suited to the present sampling rate and duration of the FID.  Therefore, the evolution of the beating signal is clearly observed on a microsecond timescale in the time domain. Each beat in FIG.~\ref{fig:fids} can be considered to be a M3WM signal from a localized state of a given chirality; thus, the observed oscillating signal originates from the field-free time-dependent coherent chiral tunneling dynamics in the target rotational state.

It is important to note that the coherent superposition of the $|\pm\rangle$ eigenstates, driven by the pump cycle, can be applied universally to all tunneling molecules, regardless of the tunneling frequency and the coherence readout scheme. For fast-tunneling chiral molecules like benzyl alcohol, the increased frequency differences between tunneling components lead to better separation of sub-cycles in the scheme, reducing the interference between them and thus improving the efficiency of coherent state preparation. However, it also complicates the observation of coherent tunneling dynamics through time-domain frequency beats, as simultaneous excitation of both probing sub-cycles becomes impractical. Alternatively, the linear phase-delay dependence can be generally applied to monitor these dynamics, as shown in FIG. \ref{fig:fft_result}a.

This study, which uncovers tunneling dynamics in a pure rotational state within the ground vibronic state, serves as a streamlined analogue to previous computational investigations of pure vibrational and rotational-vibrational model systems \cite{quack1989,vibration_jcp2002,ammonia_jcp2019}. Our results lay the groundwork for future experimental exploration involving both vibrational and rotational degrees of freedom with coherent laser fields. By employing tailored microwave fields, this coherent approach can be further extended to control the tunneling dynamics, achieving coherent inhibition and enhancement of tunneling. Through the coherent inhibition strategy, similar with coherent destruction of tunneling in laser physics \cite{cdt_prl1991}, the induced dynamically localized chiral wavepacket can be selectively confined within one of the potential wells, enabling a complete transition from delocalized $|\pm\rangle$ eigenstates to localized chiral states $|R\rangle$ or $|S\rangle$. This may open the door to resolving Hund's paradox \cite{hund1927-1,hund1927-2}. The controllable, time-dependent quantum dynamics also finds its application in examining chiral molecular systems for perturbation effects from parity violation, exploring the symmetry of space and time and its violation in chiral molecules, particularly in cases where $\Delta E_\mathrm{PV}$ is expected to be significantly larger than $\Delta E_{\pm}$ \cite{trs_jms1993,trs_chirality1998,quack_2020}.

\textit{Acknowledgments.}\textemdash This work has been funded by the Deutsche Forschungsgemeinschaft (DFG, German Research Foundation) - Projektnummer 328961117 - SFB 1319 ELCH.


\textit{Author contributions statement.}\textemdash All authors conceived the experiment, W.S. conducted the experiments, and W.S. and D.T. analyzed the results. All authors reviewed the manuscript. 

\textit{Data availability.}\textemdash All the data supporting this study's findings are available within the main text and the Supplementary Material and from the corresponding author on request.



\textit{Competing interests.}\textemdash The authors declare no competing interests.

%

\end{document}